\documentclass[showpacs,10pt,twocolumn,prl]{revtex4-1}

\usepackage{amsmath}
\usepackage{amssymb}
\usepackage{graphicx}
\usepackage{graphics}
\usepackage{epsfig}
\usepackage{color}
\usepackage{epstopdf}

\begin{document}

\title{Low-Temperature Thermopower in CoSbS}
\author{Qianheng Du$^{1,2,\dag}$, Milinda Abeykoon $^{3}$, Yu Liu,$^{1}$, G. Kotliar$^{1,4}$ and C. Petrovic$^{1,2,\ddag}$}
\affiliation{$^{1}$Condensed Matter Physics and Materials Science Department, Brookhaven
National Laboratory, Upton, New York 11973, USA\\
$^{2}$Department of Materials Science and Chemical Engineering, Stony Brook University, Stony Brook, New York 11790, USA\\
$^{3}$Photon Science Division, National Synchrotron Light Source II, Brookhaven National Laboratory, Upton, New York 11973, USA\\
$^{4}$Department of Physics and Astronomy, Rutgers University, Piscataway, New Jersey 08856, USA}

\date{\today}

\begin{abstract}
We report giant thermopower S = 2.5 mV/K in CoSbS single crystals, a material that shows strong high-temperature thermoelectric performance when doped with Ni or Se. Changes of low-temperature thermopower induced by magnetic field point to mechanism of electronic diffusion of carriers in the heavy valence band. Intrinsic magnetic susceptibility is consistent with the Kondo-Insulator-like accumulation of electronic states around the gap edges. This suggests that giant thermopower stems from temperature-dependent renormalization of the noninteracting bands and buildup of the electronic correlations on cooling.
\end{abstract}

\maketitle

Thermoelectric (TE) materials convert heat into electric power and \emph{vice versa}, which is attractive for power generation and greenhouse gas emissions reduction \cite{Snyder}. Even though efficient thermoelectrics are also needed for applications below the room temperature, progress in cryo-thermoelectric materials discovery has been modest \cite{HeJ}. In cryogenic environment there are two major differences when compared to high temperature automotive or solar applications. Since temperatures of interest are small, for high figure of merit $ZT= (S^{2}$$\sigma$/$\kappa$)$T$, where $S$ is thermopower and $\sigma$ and $\kappa$ are electrical and thermal conductivity,  thermoelectric power factor $(S^2\sigma)$ must be large. In addition, electronic correlations cannot be neglected \cite{PallsonG,KoshibaeW}. Therefore, FeSb$_{2}$-like electronic systems \cite{PetrovicC1,PetrovicC2} with largest thermopower and thermoelectric power factor  known to date \cite{BentienA,JieQ} could be essential materials for future thermoelectric cryodevices.

Thermopower of FeSb$_{2}$ reaches colossal values of up to -45 mV/K \cite{BentienA}, however, the physical mechanism is not well understood. From the Mott formula $S$=-[($\pi$$k_{B}$)$^{2}$/3e]T($\partial$ln$\sigma$/$\partial$E)$|$$_{E_{F}}$ so values of the order of $S$$\approx$-[($\pi$$^{2}$$k_{B}$)/3$e$][$T$/T$_{F}$]$\approx$10 $\mu$V/K are to be expected for diffusion mechanism in metals where states at the Fermi level take part in the conduction process \cite{CutlerM}. In semiconductors, the electronic contribution to thermpopower scales with the distance from the Fermi level in units of $k_{B}$T and with its fractional contribution to the total current. Therefore, semiconductors have typically larger diffusion thermopower when compared to metals, however commonly observed values are still in 100's of $\mu$V/K range \cite{Fritzche}. Whereas there are arguments against the phonon-drag in FeSb$_{2}$ \cite{SunP1,SunP6}, recent studies suggest nearly ballistic phonons dragging massive electrons to enhance thermopower up to $|S|$ = 27 mV/K in high-purity FeSb$_{2}$ single crystals \cite{TakahashiH}. In order to make progress and enable predictive materials design, it is important to discover new materials with high thermoelectric parameters and with tunable physical properties.

Ternary crystal structures offer higher tunability when compared to binary compounds such as marcasite FeSb$_{2}$ \cite{KangC}. Recently, it was shown that CoSbS \cite{Nahigian} could be a high-temperature thermoelectric material due to several positive factors that work simultaneously to enhance its thermoelectric performance \cite{CarliniR}. Combined linearized augmented plane wave (LAPW) theoretical and experimental studies confirmed this to be the case, unveiling additional features in the electronic structure such as substantial band degeneracy near edges, flat dispersions throughout the entire Brillouin zone and large density of states rising rapidly around the gap \cite{ParkerD}. Co$_{1-x}$Ni$_{x}$SbS and CoSbS$_{1-x}$X$_{x}$ (X=Se,Te) showed figure of merit $ZT=(0.5-0.62)$ in (730-900) K range and high power factor \cite{LiuZ,ChmielowskiR,ChmielowskiR2}. Moreover, even higher values of ZT $>$ 1 at 600 K have been predicted by density functional theory in optimized CoSbS-based materials \cite{Bhattacharya}.

Here, we unveil the low-temperature thermoelectric properties of CoSbS and report $S$ = 2.5 mV/K at 40 K. In contrast to FeSb$_{2}$ where the mechanism of colossal $S$ is still controversial, we show that such large values can be explained by the electronic diffusion in the heavy valence band. Our results imply a Kondo-insulator-like scenario of the electronic correlations buildup associated with the thermopower peak; the intrinsic magnetic susceptibility can be understood within a model of a metallic spin paramagnetism, albeit with a large low-temperature renormalization of the noninteracting bands \cite{Jaccarino,Mandrus}.

Single crystals of CoSbS were grown by heating starting materials are first at 500 $^\circ$C for 6 hours and then to 1000 $^\circ$C. Crystals were decanted at 650 $^\circ$C after the slow-cooling process. They were oriented and polished along principal crystallographic axes for four-probe resistivity and thermal transport measurements. Magnetic, transport and thermal measurements were carried out in Quantum Design PPMS-9 and MPMS-5XL. In thermoelectric measurement the magnetic field is applied along the $b$-axis whereas temperature gradient is along the $a$-axis. Synchrotron powder X-ray diffraction (XRD) data were taken at NSLS2 beamline 28-ID-1 .

\begin{figure}
\centerline{\includegraphics[scale=0.36]{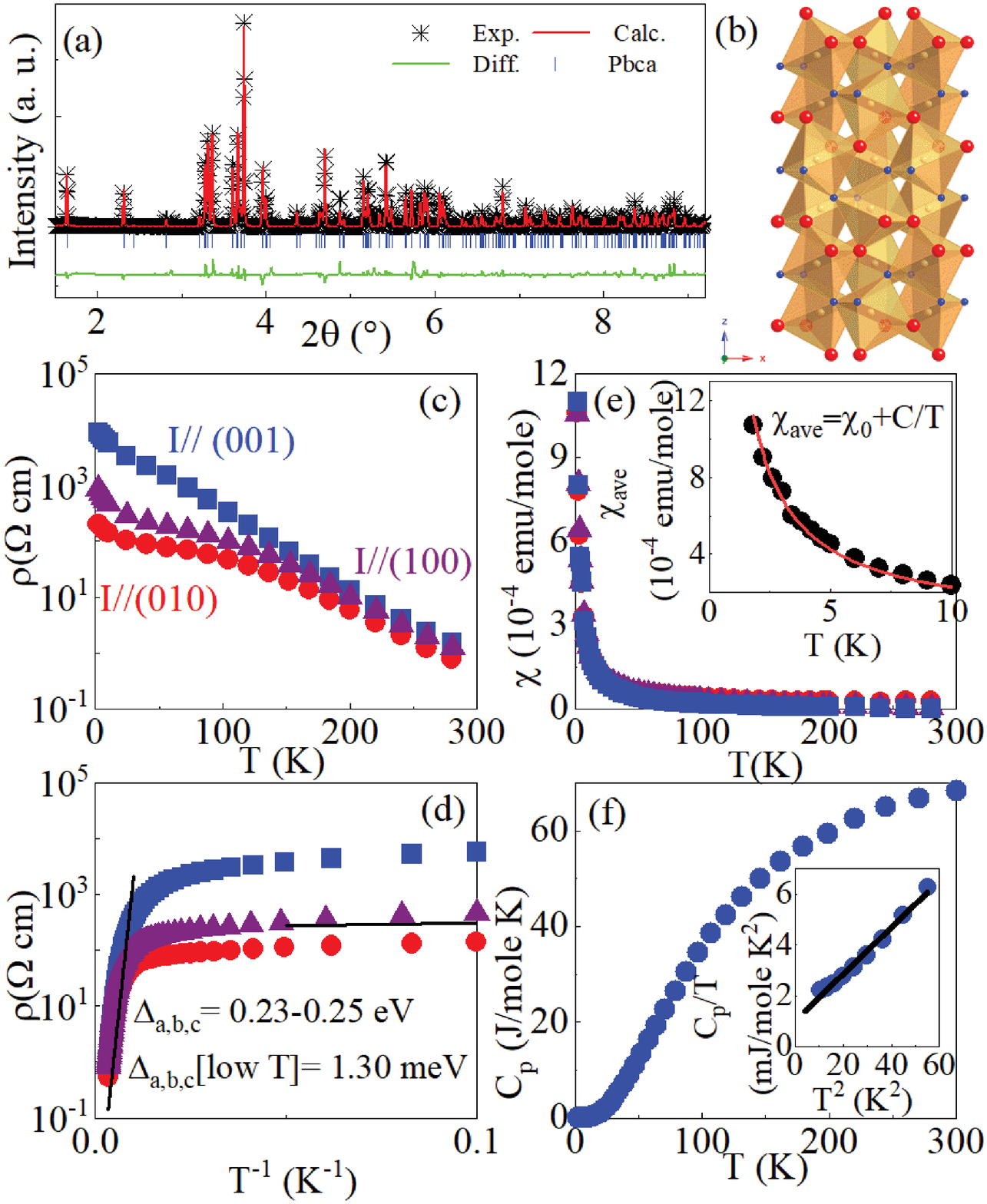}}
\caption{(Color online). (a) Powder XRD pattern and refinement. The data were shown by (+), fitting and difference curves are given by the red and green solid line, respectively. (b) Crystal structure of CoSbS. (c,d) Resistivity vs temperature for CoSbS crystals with current along different crystalline axes, revealing semiconductor behavior with different temperature regions of activated transport. (e) Anisotropic magnetic susceptibility vs temperature for a CoSbS crystal in magnetic field of 1 T. Inset shows average magnetization. (f) Heat capacity for CoSbS crystals; inset shows low temperature region.}
\label{magnetism}
\end{figure}

Powder X-ray diffraction (XRD) pattern confirms that single crystals crystallize in a \textit{Pbca} space group with refined lattice parameters a = 0.58357(1)nm, b = 0.59499(1) nm, and c = 1.16563(2) nm [Fig. 1(a)] \cite{Nahigian}. The space group is similar to pyrite and marcasite where each cation is octahedrally coordinated with six nearest neighbor anions [Fig. 1(b)]. In CoSbS, each octahedron shares an edge with one neighboring octahedron and corners with ten additional neighboring octahedra. The element analysis performed using an energy-dispersive x-ray spectroscopy (EDX) in a JEOL LSM-6500 scanning electron microscope indicated CoSbS composition.

Temperature-dependent resistivity is rather anisotropic [Fig. 1(c-d)]; $\rho_{c}$$\gg$$\rho_{a,b}$ at low temperature. There is a similar activated behavior in temperature $\rho=\rho_0exp(\Delta/2k_BT)$ over limited temperature ranges. The activated behavior in 200 $< T <$ 300 K range gives $\Delta=$ 0.23-0.25(1) eV. At low temperatures 10 K $< T <$ 20 K we detect additional small gap $\Delta_{lowT}=$ = 1.30(1) meV. Temperature dependence of the resistivity is consistent with a lightly doped semiconductor: impurity-band conduction with 1.30 meV gap dominates at low $T$ and intrinsic conduction with $\Delta$ = 0.24(2) eV dominates at high $T$ \cite{Neamen}.

CoSbS shows paramagnetic magnetic susceptibility [Fig. 1(e)], consistent with the low spin state of Co$^3$$^+$ \cite{Hulliger}. Whereas the anistropy in the $\rho(T)$ is significant, $\chi(T)$ shows much less anisotropy. The low-temperature tail could come from paramagnetic impurity effects, i.e. some residual imperfections and off-stoichiometry similar to FeSi and FeSb$_{2}$ \cite{PetrovicC1,SchlesingerZ}. We fit the average low-temperature magnetization $\chi_a$$_v$$_e$$=\frac{1}{3}(\chi_a+\chi_b+\chi_c)$ \cite{PetrovicC1} using Curie law $\chi=\chi_0+\frac{c}{T}$ [Fig. 1(e) inset] to obtain the Curie constant associated with spin 3/2 Co$^2$$^+$. The fraction of Co$^2$$^+$ is $1.6 \%$. The intrinsic impurity-free magnetic susceptibiity of CoSbS will be discussed later. Heat capacity of CoSbS crystals [Fig. 1(f)] shows the electronic specific-heat coefficient $\gamma$ from $\frac{C}{T} = \beta$$T^2 + \gamma$ is $\gamma (T\longrightarrow0) \approx 1.4(1)\times10^{-3}$ J/mole $\cdot$ K$^{2}$, as expected for a semiconductor. The Debye temperature is 426(2) K, which implies an average sound velocity of $\nu_s \approx 3700$ m/s \cite{Orson}.

\begin{figure}
\centerline{\includegraphics[scale=0.36]{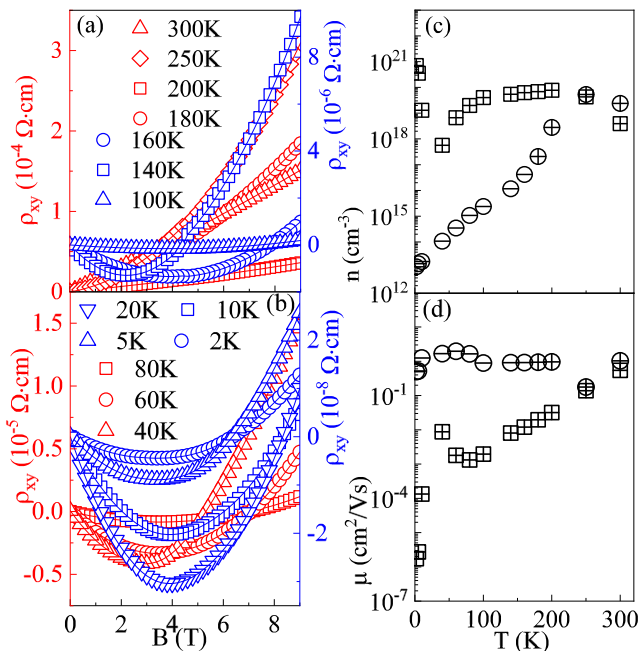}}
\caption{(Color online) (a,b) Hall resistivity ($\rho_x$$_y$) vs magnetic field at different temperatures with current along the $a$-axis. The solid lines are fits using two-bands model. (c) Carrier concentration and (d) Hall mobility as a function of temperature. + denotes the hole carriers and - denotes the electron carriers. Squares represent carrier 1 and circles represent carrier 2, as delineated in the text.}
\label{magnetism}
\end{figure}

\begin{figure}
\centerline{\includegraphics[scale=0.42]{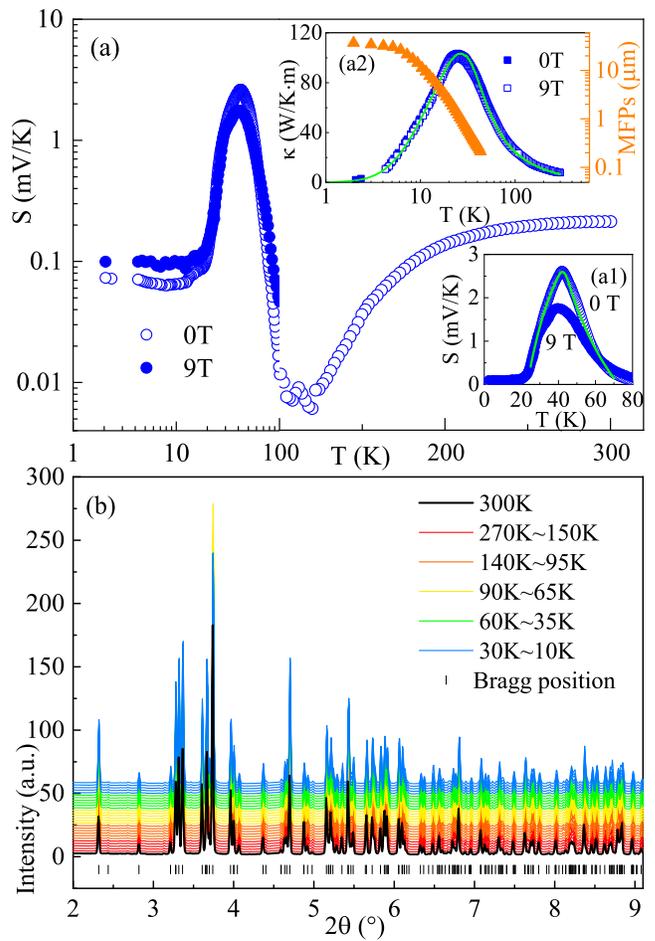}}
\caption{(Color online) (a) Thermopower of CoSbS crystal with the temperature gradient along (100) measured below 300 K. (b) Powder XRD patterns from room temperature to 10 K taken in 10 K steps below 90 K and 30 K steps above 90 K. There are no new reflections in the temperature range of thermopower rise, indicating the absence of the structural transition. Insets: (a1) Low-temperature thermopower and fit using single-band model (see text) (a2) Thermal conductivity for the same crystal and the corresponding phonon mean free path. The green line shows the fitting of thermal conductivity.}
\label{magnetism}
\end{figure}

Hall resistivity $\rho_{xy}$ in magnetic field [Fig. 2(a,b)] is not linear, confirming the presence of multiband electronic transport \cite{ParkerD}. In a two-carrier system, the Hall coefficient is $R_H = \rho_x$$_y$$/H = \rho_0(\alpha_2 + \beta_2H^2)/(1 + \beta_3H^2)$, where $\alpha_2 = f_1\mu_1 + f_2\mu_2$, $\beta_2 = (f_1\mu_2 + f_2\mu_1)\mu_1\mu_2$, and $\beta_3 = (f_1\mu_2 + f_2\mu_1)^2$, where $\rho_0$ = $\rho($$\mu_0H = 0)$, $f_i = |n_i\mu_i|/\Sigma|n_i\mu_i|$ is the $f$ factor, and $n_i$ and $\mu_i$ are individual carrier band concentrations and mobilities \cite{Kim2}. From the model we obtain both carrier band concentrations and mobilities [Fig. 2(c,d)]. Different from the conventional electrons or holes that are associated with a continuous energy band, "carrier" denotes a set of carriers with identical mobility corresponding to only one energy and/or one degenerate energy level in the model. Carriers in the low mobility band are denoted as carrier 1 and in the high mobility band are carrier 2. In the high temperature range where the intrinsic conduction dominates ($T>$ 180 K), both the carrier 1 and carrier 2 are hole-like whereas in the temperature range where extrinsic conduction dominates, the carrier 1 is hole-like and the carrier 2 is electron-like. At 300 K, the concentration of carrier 2 ($n_2 \sim 10^{19}$$cm$$^{-3}$) is about one order of magnitude larger than carrier 1 ($n_1 \sim 10^{18}$$cm$$^{-3}$). Room-temperature carrier concentration values are similar to the values observed in narrow-gap Kondo semiconductors FeGa$_{3}$ and Ce$_{3}$Bi$_{4}$Pt$_{3}$ \cite{HadanoY,PietrusT}. There is a sharp decrease of the carrier 2 concentration below 200 K to $n_2 \sim 10^{13}$ at 2 K. At the same time, the $n_1$ increases about one order of magnitude, keeping the total carrier concentration unchanged. Another difference between the two carrier types is in the Hall mobility. The mobility of carrier 2 ($\mu_2$) is one to several orders larger than mobility of carrier 1 ($\mu_1$) below 250 K. Above 200 K the intrinsic conduction dominates and the conduction of the accepter band occurs with low mobility. The carrier 2 comes from the intrinsic carriers and extrinsic carriers in the valence band and the carrier 1 is extrinsic carrier in the impurity band. The crossover of $n_1$ and $n_2$ could possibly be explained by the freezing of extrinsic carriers into the accepter band about 0.5 meV above the valence-band edge [Fig. 1(c,d)].

Thermopower $S(T)$ shows a peak at 40 K [Gig. 3(a)]; its sign is consistent with low-temperature Hall data below 300 K and the curve is very similar to that of FeSb$_2$ crystals \cite{BentienA,TakahashiH}. Having in mind the dominant carrier type in the temperature region of thermopower enhancement, we consider single-degenerate parabolic band: \cite{SunP1,BehniaK}\\
\begin{equation}
   \centering
   S(T) = \pm \frac{\pi^2}{3}\frac{k_B}{e}\frac{k_BT}{\epsilon_F}(r + \frac{3}{2})
\end{equation}
Within the free-electron approximation, the Fermi energy $\epsilon_F$ is $\epsilon_F=h^2/2m^\star(3n/8\pi)^{2/3}$ whereas the $m^\star=m_0$, where $m_{0}$ is the free-electron mass, and $n$ could be estimated from the Hall effect. The carrier scattering parameter is usually between -0.5 and 1.5. Here it is assumed to be $-\frac{1}{2}$; however fit results do not critically depend on the $r$ value. Notably, to describe the observed peak in $S(T)$, an enhancement factor 6.5 in the $m^\star$ from free electron mass is required. This points to the strong electronic correlations and a relatively  heavy valence band. The fitted line is shown in Fig. 3(a) as a green solid line. We can exclude phase transition contribution to thermopower since CoSbS shows neither metal-insulator [Fig. 1(c)] nor crystal structure change [Fig. 3(b)] at 40 K. Thermopower peak arising from spin-fluctuation mechanism in semiconductors occurs at temperatures of the long-range or the short-range magnetic order \cite{TsujiiN,MentreO}. Since CoSbS does not show any magnetic order, spin-fluctuation mechanism of thermopower enhancement is unlikely.

The lattice thermal conductivity is usually treated using Debye approximation \cite{Callaway}:\\
\begin{equation}
   \centering
   K_L = \frac{k_B}{2\pi^2\nu_s}(\frac{k_B}{\hbar})^3T^3\int_{0}^{\frac{\theta_D}{T}}\frac{\tau_cx^4e^x}{(e^x-1)^2}dx
\end{equation}
\\where $x=\frac{\hbar\omega}{k_BT}$ is dimensionless, $\omega$ is the phonon frequency, $k_B$ is the Boltzmann constant, $\hbar$ is the Plank constant, $\theta_D$ is the Debye temperature, $\nu_s$ is the  velocity of sound and $\tau_c$ is the relaxation time. The overall relaxation rate $\tau_c^{-1}$ can be determined by combining different scattering processes\\
\begin{equation}
\begin{aligned}
  \tau_c^{-1} &= \tau_B^{-1}+\tau_D^{-1}+\tau_U^{-1}\\
              &= \frac{\nu}{L}+A\omega^4+B\omega^2Te^{-\frac{\theta_D}{T}}
\end{aligned}
\end{equation}
\\where $\tau_B$, $\tau_D$, $\tau_U$ are the relaxation times for boundary scattering, defect scattering, and Umklapp processes, respectively. The adopted values of the parameters are $A=3.0\times10^{-43}$ $s^3$, $B=6.2\times10^{-18}$ $sK^{-1}$ and the phonon mean free path at low temperature in the boundary scattering regime is $L=6.6\times10^{-5}$ $m$. It should be noted [inset (a2) in Fig. 3(a)] that $\kappa(T)$ values are smaller when compared to FeSb$_{2}$, probably due to enhanced defect and Umklapp scattering processes of high-momentum phonons \cite{BentienA}.

Next, we discuss the mechanism of giant thermopower in detail. It is instructive to evaluate the contribution of the phonon drag. First, we note that the electronic contribution to $\kappa(T)$ using Wiedemann-Franz law is negligible; i.e. phonon $\kappa_{P}(T)$ and total $\kappa(T)$ are indistinguishable. Phonon drag thermopower is  directly related to phonon mean free path $l_{p}$ by $S_{ph} = \beta\nu_sl_p/\mu$$T$ \cite{Herring,Weber}, where $\beta$ parameter describes the relative interaction strength of electron (or hole) and phonon with $0 < \beta < 1$, and $\mu$ is the mobility of hole in the valence band. Since $\kappa$ $=\frac{C_v\nu_s l_p}{3}$ \cite{Goldsmid}, we estimate phonon mean free path $l_p$ to be about 0.7 $\mu$m at the maximum thermopower and 53 $\mu$m at 2 K [Fig. 3(a) inset (a2)], consistent with Callaway model. This is several orders of magnitude smaller when compared to not only FeSb$_{2}$ but also to typical materials with phonon-drag thermopower mechanism such as Ge \cite{SunP1,NguyenL}. Together with the observation that $S(T)$ and $\kappa(T)$ are maximized at different temperatures [Fig. 3 inset (a,b)], present data strongly suggest that phonon-drag mechanism is not dominant in CoSbS. Furthermore, the maximum $S$ value decreases from $2.5$ mV/K in 0 T to $1.7$ mV/K in 9 T [Fig. 3 inset (b)]. This confirms the importance of electron diffusion in giant low-temperature thermopower in CoSbS and calls for the investigation of the m$^{*}$ enhancement origin.

\begin{figure}
\centerline{\includegraphics[scale=0.4]{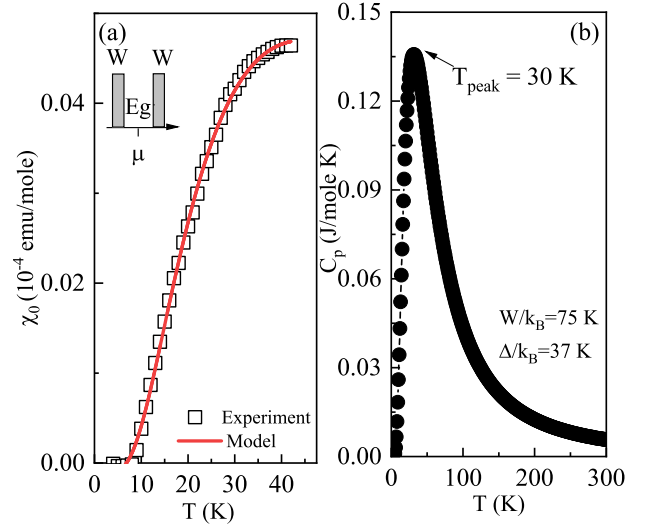}}
\caption{(Color online)  (a) Intrinsic low-temperature $\chi$(T) obtained by subtraction of the Co$^{2+}$ impurity Curie-Weiss tail. The red line shows the fitting by narrow-band-small-gap model (see text). The fitting parameters are listed. (b) The contribution to heat capacity from spin transition calculated by parameters obtained from magnetic susceptibility.}
\label{magnetism}
\end{figure}

By subtracting impurity susceptibility from the measured $\chi(T)$ we obtain the intrinsic magnetic susceptibility [Fig. 4(a)]. There is a rise of paramagnetic spin susceptibility with temperature increase that can be explained with a narrow-band-small-gap model \cite{Jaccarino} with a pileup of states in the two narrow bands of width $W$, separated by an energy gap $E_g=2\Delta$ [inset in Fig. 4(a)]. This model has been successfully applied to point to Kondo-insulator-like physics in correlated electron semiconductors FeSi and FeSb$_2$ \cite{PetrovicC2,Mandrus}.

The Pauli susceptibility of an itinerant electron system with $N(E)$ density of states is\\
\begin{equation}
   \centering
   \chi(T) =-2\mu_B^2\int_{c-band}{N(E)\frac{\partial{f(E,\mu,T)}}{\partial{E}}}dE,
\end{equation}
\\where $\mu_B$ is the Bohe magneton and $f(E,\mu,T)=(exp[(E-\mu)/k_BT]+1)^{-1}$. The factor 2 in the equation is due to the holes in the valence band that contribute to $\chi$ in the same way as electrons in the conduction band. Taking $N(E)=Np/W$ and $\mu=W+\Delta$, where $p$ is the number of states/cell, and N is number of unit cells, we obtain\\
\begin{equation}
   \centering
   \chi(T) =-2\mu_B^2\frac{Np}{W}\frac{exp(\beta\Delta)(1-exp(\beta\Delta))}{(1+exp(\beta\Delta))(1+exp(\beta(\Delta+W)))}+\chi_c.
\end{equation}
\\By fitting magnetic susceptibility data [Fig. 4(a)], we obtain: $W/k_B$ = 75 K and $\Delta/k_B$ = 37 K. Using the same parameters obtained from the magnetization fit, we have calculated spin state transition contribution to specific heat $C_P=(\partial{U}/\partial{T})$ with $U$ given by the contribution from valence and conduction band\\
\begin{equation}
\begin{aligned}
   U =&\int_{0}^{W}\frac{Np}{W}\frac{EdE}{exp(\beta(E-W-\Delta))+1}\\
   &+\int_{W+2\Delta}^{2W+2\Delta}\frac{Np}{W}\frac{EdE}{exp(\beta(E-W-\Delta))+1}.
\end{aligned}
\end{equation}

The calculated heat capacity [Fig. 4(b)] shows a clear Schottky peak around 30 K, which is expected for a two levels electronic system separated by narrow gap, however the phonon contribution is dominant in the measured temperature range [Fig. 1(f), Fig. 4(b)], i.e. the phonon heat capacity dwarfs the Schottky peak contribution. We note that the difference between low-temperature energy gaps extracted from $\rho(T)$ and $\chi(T)$ could be explained by invoking the existence of the smallest indirect energy gap relevant for low-temperature transport, but not for $\chi$ \cite{PaschenS}.  Hence, magnetic properties indicate that the giant increase in low-temperature thermopower comes from the Kondo-insulator-like low-temperature distortion in the density of the electronic states \cite{Heremans}. This is further supported by the rather low carrier mobility [Fig. 2(d)], comparable to values found in rare earth hybridization compounds and FeSi  \cite{SalesB,FiskZ}. The low mobility of carriers implies high effective band mass and calls for detailed photoemission studies.

In mixed anion compounds FeSbP and FeAsP related to FeSb$_{2}$, the size of the band gap is related to a rotation angle $\theta$ of the Fe-centered octahedron around the $z$ axis \cite{KangC}. Since the Co(Sb,S)$_{6}$ octahedra exhibit relatively large rotation displacements when compared to Fe(Sb,P/As)$_{6}$ due to corner-sharing, larger intrinsic band gap could be favored by the local crystallographic arrangement. The thermal conductivity of CoSbS is large and electrical conductivity is much too low for a competitive thermoelectric material. However, since transition metal character dominates near the band edge in DFT calculations, alloying on Sb or S atomic site may improve electrical conductivity and may also be more effective in thermal conductivity reduction when compared to bulk nanostructuring methods due to very low sound velocity anisotropy \cite{ParkerD}.

\textit{Summary} - We have evaluated low-temperature thermoelectric properties of CoSbS, a material that shows high thermoelectric performance above the room temperature. Here we show that CoSbS single crystals also exhibit giant thermopower at cryogenic temperatures. The mechanism of thermopower enhancement involves buildup of the electronic correlations on cooling and consequently giant thermopower could be explained within the model of diffusive correlated electrons. Our work calls for further investigation of the Kondo-insulator-like electronic correlations that develop at low temperature and for the relation of the diffusive thermopower to the 0.50 meV gap absent in DFT calculations \cite{ParkerD}. The question of how giant diffusion thermopower in CoSbS is related to electronic structure and impurity states is an interesting subject for future study.

Work at Brookhaven is supported by the Research supported by the U.S. Department of Energy, Office of Basic Energy Sciences as part of the Computation Material Science Program (Y.L., G. K. and C.P.). This research used 28-ID-1 (PDF) beamline of the National Synchrotron Light Source II, a U.S. Department of Energy (DOE) Office of Science User Facility operated for the DOE Office of Science by Brookhaven National Laboratory under Contract No. DE-SC0012704.

$\ddag$petrovic@bnl.gov
$\dag$qdu@bnl.gov

\end{document}